# Temperature Control of Electromigration to form Gold Nanogap Junctions

G. Esen, M.S. Fuhrer[a]

Department of Physics and Center for Superconductivity Research, University of

Maryland, College Park, MD 20742-4111, USA

Controlled electromigration of gold nanowires of different cross-sectional areas to form nanogap junctions is studied using a feedback method. A linear correlation between the cross sectional area of the gold nanowires and the power dissipated in the junction during electromigration is observed, indicating that the feedback mechanism operates primarily by controlling the temperature of the junction during electromigration.  We also show that the role of the external feedback circuit is to prevent thermal runaway; minimization of series resistance allows controlled electromigration to a significant range of junction resistances with a simple voltage ramp.

a) Author to whom correspondence should be addressed.  Electronic mail: mfuhrer@physics.umd.edu

Electromigration (EM), the electrical current-induced diffusion of atoms in a thin metal film, is important as a failure mode in integrated circuit interconnects[1]. Failure of a narrow metal wire due to EM has also recently been utilized extensively to prepare stable electrical contact pairs with nanometer separation for single molecule electrical experiments [2,3]. It has been suggested [2-4] that the dominant failure mechanism in such electrically stressed gold nanowires is thermally-assisted EM. Recently, feedback schemes for controlling the rate of EM were shown to allow control over the final junction resistance (and presumably gap size) in such nanogap junctions [5,6].

The empirical formula for the median time to failure $MTTF = Aj^{-n}e^{\frac{E_a}{kT}}$, known as the Black equation[7] has been employed to determine interconnect reliability, where $A$ is a sample dependent constant, $j$ the current density, $n$ an exponent empirically found between 1 and 7 [8], $E_a$ an activation energy for atomic motion, $k$ the Boltzmann constant, and $T$ the temperature. However, in the formation of nanogap junctions by EM of a short nanowire, the current and temperature (due to Joule heating) are both changing rapidly. Understanding the role of current and temperature is then critical to design of circuits to produce nanogap junctions in a controlled manner.

In recent works[5,6] it is proposed that controllable electromigration occurs at constant applied power. Furthermore it is stated[5] that electromigration is triggered at a constant temperature. In this letter, by experimenting on different nanowire geometries, we confirm that electromigration occurs at constant temperature independent of geometry, and show that the feedback mechanism works primarily through controlling the temperature of the electromigrating junction. Furthermore, we estimate the temperature of the junction during EM to be only a few hundred Kelvins, low enough to allow study of many molecular adsorbates without desorption or dissociation of the molecules. We also show that the role of the external feedback circuit is to prevent thermal runaway in the junction, allowing control in the region in which the temperature increases with increasing junction resistance. This region can be minimized by minimizing series resistance in the circuit, allowing significant control of EM with a simple voltage ramp.

Our devices are fabricated in two steps using conventional electron beam lithography (EBL) and lift-off on $SiO_2$ (500 nm)/Si substrates. We first fabricate thin gold lines (with no adhesion layer) of width 40 nm to 100 nm and thickness 15 to 30 nm. In the second EBL step, we deposited 5nm Cr and 70 nm Au to form contacts and bonding pads. A scanning electron microscope (SEM) micrograph of such a device is shown in the inset of Figure 1.

To control the EM process we used a computer-controlled feedback scheme similar to ref. [5], consisting of the following steps: We first measure a reference conductance value at a voltage of 100 mV. We then increase the voltage until the conductance drops by a set fraction (typically 2-5%) of the reference conductance value. At this point the voltage bias is ramped down by 50 to 100 mV (at a rate of 50 mV/s) and a new reference conductance value is measured. We repeat this process until the desired conductance is reached. We performed our experiments in a gas flow $^4$He cryostat at substrate temperatures from 1.3 K to room temperature (RT).

A representative current vs. voltage ($I$-$V_{bias}$) curve taken through a feedback-controlled EM process for one of the devices is shown in Figure 1. The data labeled A shows a smooth $I$ vs. $V_{bias}$ curve indicating that EM has not begun in the gold wire.

Although there is a resistance increase with increasing bias in A, we found that if we stop the voltage bias in region A, this resistance increase is reversible. Such a reversible resistance increase shows that the gold wire heats up before EM begins. The data labeled B show that after this initial heating the gold wire begins to change resistance irreversibly due to EM. The data labeled C show that one can stop and restart the voltage bias before the gold wire totally fails. $I$-$V_{bias}$ curves of two bias processes perfectly match each other indicating that in the second biasing process the gold wire first heats up to the temperature where significant EM takes place, and then EM restarts.

We now discuss the temperature of the wire during EM. We assume the total resistance of the circuit $R$ is the sum of two resistors, $R_L$, the lead resistance (equal to the total measured resistance at low bias) and $R_J$, the resistance of the "junction", the weak spot formed in the wire by EM; i.e. $R_J = 0$ initially. The power dissipated in the junction is then $P_J = I^2 R_J$. Note that $R_J$ includes a contribution from the resistance change of the leads upon heating; and heating due to the resistance of the nanowire itself is ignored. Hence $P_J$ is a rough estimate of the power that is heating the junction, but should be valid when $R_J$ is significantly non-zero.

Figure 2 shows $P_J$ vs. $V_J = IR_J$ where the inset graph is the corresponding $I$-$V_{bias}$ curve of the data. After the junction begins to increase resistance due to EM [point (a)] the power dissipated in the junction reaches a relatively constant value. Constant power dissipation in the junction is observed over an order of magnitude of junction voltage.

Figure 3 shows that the power dissipated in the junction during electromigration is proportional to the nanowire area. Since the thermal conductance of the wire is proportional to the cross-sectional area, the power required to maintain a given wire temperature should also be proportional to area. Thus we conclude from Figure 3, and the relatively constant power observed in Figure 2, that the electromigration rate is dominated by temperature, and the feedback scheme operates to control the wire temperature. This conclusion is reasonable from the Black equation, given that temperature enters exponentially, and voltage (through current density) only algebraically.

We now estimate the temperature of the junction $T_J$ during EM. Here we neglect the heat conduction to $SiO_2$ substrate by considering the relative magnitude of the thermal conductivity of gold and $SiO_2$ [9] and consider the contacts as infinite heat sinks at $T = 1.5 K$. The temperature at the midpoint of a wire with uniform power generation over its volume is $T_J = \frac{PL^2}{8V\kappa}$ where $P, V, L, \kappa$ are respectively the total power generated in the nanowire, the volume and length of nanowire and the thermal conductivity of gold [10]. For the wire in Figure 2, the maximum power is estimated as $\sim 0.67 mW$ (including $P_J$ at point (a) and additional power generated due to the estimated resistance 11 Ω of the nanowire at $T$ = 1.5 K); using the thermal conductivity of gold as $\geq 320 \frac{W}{mK}$ (the room-temperature value for bulk gold) we estimate $T_J \leq 145$ K. If we consider instead that all the power is being generated at the center of the nanowire and carried out to the leads by the nanowire, then $T_J = \frac{PL^2}{4V\kappa} \leq 290$ K; which is still low enough to allow study of many molecular adsorbates without desorption or dissociation of the molecules.

We now discuss why the feedback process is feasible. The thermal time constant[11] $\tau_{th} = L^2 \rho C_p / \pi^2 \kappa$, where $C_p$ is the specific heat, and $\rho$ the density, is less than 1 ns in our wires; much faster than our external feedback circuit. This suggests that the electromigration process itself must occur very slowly. If this is the case, is feedback needed at all? I.e., could $V_{bias}$ simply be turned to zero at the desired $R$? The answer is no – at constant junction temperature (i.e. following the $I$-$V_{bias}$ curve in e.g. Figure 1) $I$, and hence $R$, is multiple-valued at a given $V_{bias}$. Stated another way, the change in temperature with junction resistance $dT_J/dR_J$ during EM must be negative to prevent thermal runaway once EM begins. Assuming that $T_J$ is proportional to $I^2R_J$, for our simple series circuit model $dT_J/dR_J < 0$ implies $R_J > R_L$, corresponding to the stable branch of the $I$-$V_{bias}$ curve where $dV_{bias}/dR > 0$ during EM (the voltage increases as electromigration progresses).

Figure 4 illustrates this instability. We turned off the feedback at various points during the EM process, solely ramping the voltage upwards at a fixed rate. The red curve shows the feedback turned off while on the stable $I$-$V_{bias}$ branch (positive $dV_{bias}/dR$); the current decreases smoothly with increasing voltage from this point. However, when the feedback is turned off on the unstable $I$-$V_{bias}$ branch (negative $dV_{bias}/dR$; blue and black curves), the current drops rapidly to the stable branch at the same $V_{bias}$. Thus the feedback scheme is only necessary to produce final resistances $R_J < R_L$; with suitable circuit design (minimization of $R_L$; i.e. short nanowires with highly conducting leads) small final $R_J$'s may be produced using a simple voltage ramp. Note that in some circuits (e.g. Figure 1) the stable $I$-$V_{bias}$ branch does not extend beyond the unstable branch; in such cases a simple voltage ramp causes abrupt failure of the wire by melting (as observed via post-mortem SEM), resulting in large (> 10 nm) gaps.

In conclusion we performed controllable electromigration on nanowires with different cross sections. We found that the average power dissipated in the junction during EM increases linearly with the area of the junction indicating the temperature control of the process and confirming that the mechanism is thermally-assisted EM. Using the maximum power dissipated in a typical device, we estimate the junction temperature during EM performed at $T = 1.5K$ to be only a few hundred Kelvins. We also note that the role of the feedback process in controlling EM is to prevent thermal runaway in the region of positive d$T_J$/d$R_J$. This region can be reduced by reducing the series resistance in the circuit, allowing controlled EM with a simple voltage ramp.

This work has been supported by the U.S. Department of Energy under Grant No. DE-FG02-01ER45939 and by the NSF-Materials Research Science and Engineering Center under Grant No. DMR-00-80008.

Figure Captions

**Figure 1** : Current vs. bias voltage during the feedback-controlled electromigration of an Au wire at T=1.3 K. Part A is a smooth curve indicating than the EM has not begun whereas in part B the resistance of the line increases irreversibly due to EM. Both parts A and B are recorded in a single voltage biasing process, producing a final resistance of ~120 Ω. At this point the voltage was reduced to zero for some time. When the bias process was restarted in C, the wire resistance is the same, demonstrating that the EM process may be frozen by turning off the voltage. The inset shows the SEM micrograph of one of our devices. The scale bar in the inset is 2 micrometers long. Arrows indicate the progression of the curve.

**Figure 2**: Power dissipated in the junction $P_J$ vs. the voltage drop at the junction $V_J$ (quantities defined in text). The irreversible change in resistance due to electromigration starts at the point labeled (a). Inset shows the corresponding current vs. bias voltage data. The starting nanowire has dimensions 830 nm long x 60 nm wide x 25 nm thick; the length and width of the nanowire is determined using SEM, and thickness by quartz crystal monitor during gold film deposition.

**Figure 3:** Power dissipated in the junction during electromigration vs. nanowire cross-sectional area. The power is the average power in the region of near-constant power seen in Figure 2.

**Figure 4:** Current vs. bias voltage during the electromigration of three similar gold wires (600-700 nm long x 40 nm wide x 15 nm thick) at T=1.3 K. For the blue and red curves, the external feedback is turned off at the points marked by the blue and red arrows respectively. For the black curve, no external feedback was used.

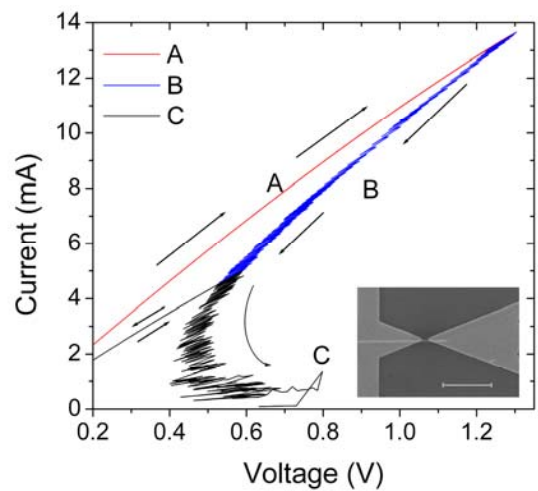

Figure 1.
G. Esen
Applied Physics Letters

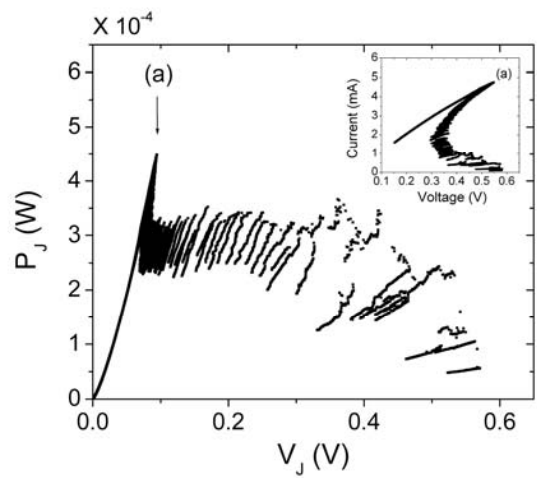

Figure 2.
G. Esen
Applied Physics Letters

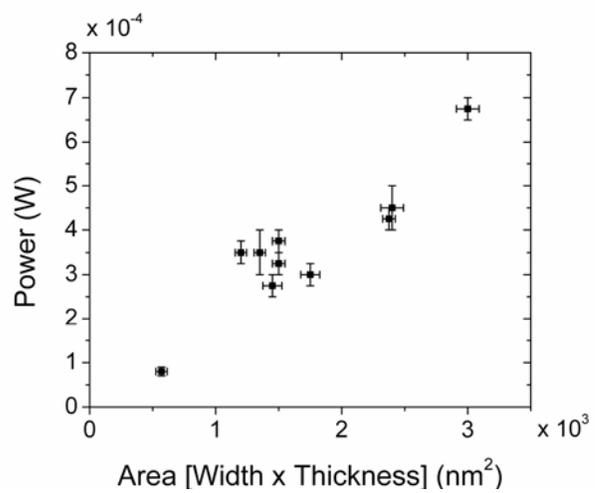

Figure 3.
G. Esen
Applied Physics Letters

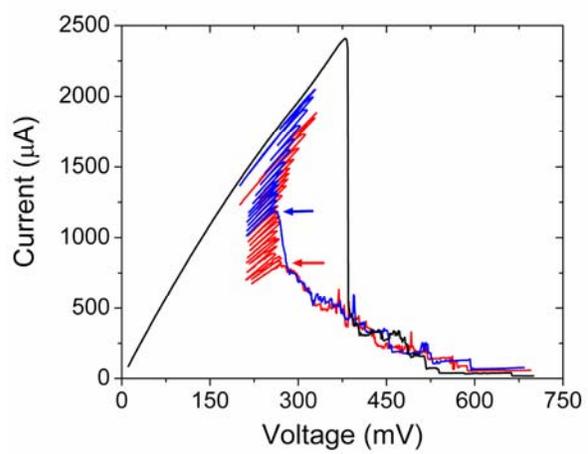

Figure 4.
G. Esen
Applied Physics Letters